\documentclass[letterpaper,10pt]{article}
\usepackage{osameet2}

\usepackage{amsmath,amssymb,amsthm,amsfonts}
\usepackage{graphicx}

\usepackage[pdftex,colorlinks=true,bookmarks=false,citecolor=blue,urlcolor=blue]{hyperref} 

\begin{document}

\title{Nonlinearity-tolerant 8D modulation formats by set-partitioning PDM-QPSK}

\author{D. F. Bendimerad$^*$, H. Hafermann, and H. Zhang}
\address{Mathematical and Algorithmic Sciences Lab, Paris Research Center, Huawei Technologies France SASU}
\email{$^*$djalal.falih.bendimerad@huawei.com}

\begin{abstract}
	We present two new nonlinearity tolerant modulation formats at spectral efficiencies lower than 4bits/4D-symbol, obtained using a simplified bit-to-symbol mapping approach to set-partition PDM-QPSK in 8 dimensions.
\end{abstract}


\section{Introduction}
In recent years and after the standardization of coherent optical transmission systems, phase modulation formats have been widely investigated \cite{Karlsson2009,Kojima2017}. Different studies concluded that designing high-dimensional modulation formats can increase the product of reach and capacity of the transmission system by either increasing the Euclidean distance, or mitigating fiber nonlinear impairments \cite{Eriksson2014b,Reimer2016,Millar2014a}.
These formats also allow an increase in the spectral efficiency granularity \cite{Reimer2016,Kojima2017}.

For instance, Shiner et al. designed an eight-dimensional (8D) nonlinearity-tolerant modulation format using the polarization-balance concept at a spectral efficiency of 2bits/4D-symbol \cite{Shiner2014}. They experimentally demonstrated a 1dB gain in net system margin with comparison to Polarization-Division-Multiplexed Binary-Phase-Shift-Keying (PDM-BPSK) standard format. Later, Reimer et al. used the same concept to design a modulation format with a higher spectral efficiency of 3bits/4D-symbol \cite{Reimer2016}.


In this paper, we propose a unified approach to construct nonlinearity-tolerant modulation formats in the spectral efficiency range of 2-4bits/4D-symbol by set-partitioning Polarization-Division-Multiplexed Quadrature-Phase-Shift-Keying (PDM-QPSK) in 8D. The aforementioned modulation formats can be seen as special cases of this approach.
We construct two new 8D formats at spectral efficiencies of 2.5 and 3.5bits/4D-symbol. Using numerical simulations we demonstrate an increased nonlinearity tolerance and distance-capacity product compared to standard formats.

\section{Design of 8D modulation formats based on set-partitioned PDM-QSPK}
PDM-QPSK symbols exhibit four distinct States of Polarization (SOPs) and a constant modulus. In eight dimensions (considering two consecutive time slots $T_1$ and $T_2$), PDM-QPSK symbols can be grouped into three disjoint sets according to the relative orientation of their SOPs in consecutive time slots:
$^{(i)}$ $64$ symbols with opposite SOPs in $T_1$ and $T_2$ (Polarization Balanced, PB);
$^{(ii)}$ $128$ symbols with orthogonal SOPs (Polarization Alternating, PA); 
$^{(iii)}$ and $64$ symbols with identical SOPs (Polarization Identical, PI).

We obtain 8D modulation formats with a high nonlinearity tolerance by selecting symbols from the three sets until the desired spectral efficiency is attained, while prioritizing PB over PA over PI symbols.
Symbols are further selected such that $^{(a)}$ the minimum Euclidean distance is maximized and $^{(b)}$ the constellation is symmetrical: each constellation point has the same number of neighbors at a given Euclidean distance.

\section{PB-5B8D and PA-7B8D modulation formats}
The $2^8=256$ symbols of PDM-QPSK in 8D are labeled by $8$ bits ($b_1,~b_2,~...,~b_8$ with $b_1$ being the most significant bit), where the first four bits determine the constellation point of a Gray-mapped standard PDM-QPSK in $T_1$, and the last four the one in $T_2$.

We obtain a new modulation format at a spectral efficiency of 2.5~bits/4D-symbol, by selecting $2^5=32$ from 64 PB symbols of set $^{(i)}$. They are labeled using $5$ information bits and 3 overhead bits in 8D. We therefore refer to this format as polarization-balanced 5 bits in 8D (PB-5B8D).
The three overhead bits, whose values are determined by choosing symbols from the set $^{(i)}$ according to the constraints $^{(a)}$ and $^{(b)}$, are obtained through the following formulas:
\begin{equation}
\label{equation:PA-5B8D_parity_bit}
b_6~=~b_3\oplus \left( b_4\oplus b_5\right)~~~~;~~~~
b_7~=~\overline{b_2}\oplus \left( b_4\oplus b_5\right)~~~~;~~~~
b_8~=~\overline{b_1}\oplus \left( b_4\oplus b_5\right).
\end{equation}
Here, ``$\oplus$" denotes the logical XOR and an overline indicates negation.

A spectral efficiency of 3.5bits/4D-symbol or $7$ information bits ($b_1~...~b_7$) in 8D requires $2^7=128$ symbols, which exhausts the available PB symbols. We add $64$ PA symbols and correspondingly refer to this format as polarization alternating 7 bits in 8D (PA-7B8D).
Here, selecting the symbols from the sets $^{(i)}$ and $^{(ii)}$ subject to constraints $^{(a)}$ and $^{(b)}$ translates into a single condition on the overhead $b_8$, which is computed as follows,
\label{equation:PA-7B8D_parity_bit}
\begin{align}
	b_8={} & \overline{b_1\oplus b_4\oplus b_6\oplus \left( b_1 \cdot b_3\right) \oplus \left( b_1 \cdot b_4\right) \oplus \left( b_1 \cdot b_5\right) \oplus \left( b_1 \cdot b_6\right) \oplus \left( b_2 \cdot b_3\right) \oplus} \\
	& \overline{\left( b_2 \cdot b_4\right) \oplus \left( b_2 \cdot b_5\right) \oplus \left( b_2 \cdot b_6\right) \oplus \left( b_3 \cdot b_5\right) \oplus \left( b_3 \cdot b_6\right) \oplus \left( b_4 \cdot b_5\right) \oplus \left( b_4 \cdot b_6\right)},
\end{align}
where ``$\cdot$" denotes the logical AND operation.

The above formulas are examples of nonlinear coding reflecting nonlinear constraints on the symbols.
We emphasize that the PDM-QPSK devices and components can be reused in implementing these formats. Complexity is increased at the transmitter because the overhead bits have to be computed according to eq.\eqref{equation:PA-5B8D_parity_bit} or eq.\eqref{equation:PA-7B8D_parity_bit}, respectively.

\section{Linear and Nonlinear channel performance}
The simulation setup is as follows: five PDM and Wavelength Division Multiplexed (WDM) channels are generated at the transmitter, where a binary sequence of length $2^{16}$ is carried in each polarization. Signals are sampled at 64samples/symbol, and modulated at a baudrate of 32Gbaud with a Root-Raised Cosine (RRC) pulse shape with a roll-off factor of 0.1. After fixing the center channel at the wavelength 1550nm and multiplexing all channels on a 37.5GHz grid, the resulting optical signal at the output of the multiplexer is launched into the optical link. The latter consists of several spans of a 75km long Large Effective Area Fiber (LEAF), an ideal 100\% in-line chromatic Dispersion Compensation Fiber (DCF) and a noiseless flat-gain in-line amplifier. The Polarization Mode Dispersion (PMD) of the fiber is not considered. In front of the receiver, White Gaussian Noise (WGN) is loaded to the signal assuming a Noise Figure of the amplifiers NF=7dB. At the receiver, an RRC matched filter is applied to the digitized signal of the center channel, followed by an 11-tap butterfly-FIR equalizer trained by 1024 symbols, which recovers both polarizations without any phase ambiguity. A Maximum Likelihood (ML) decision is performed in 8D. Finally the $Q^2$ factor is computed using a Monte Carlo loop while counting at least 400 errors.\\
\begin{figure*}[!!!ht]
	\centering
	\includegraphics[scale=0.85]{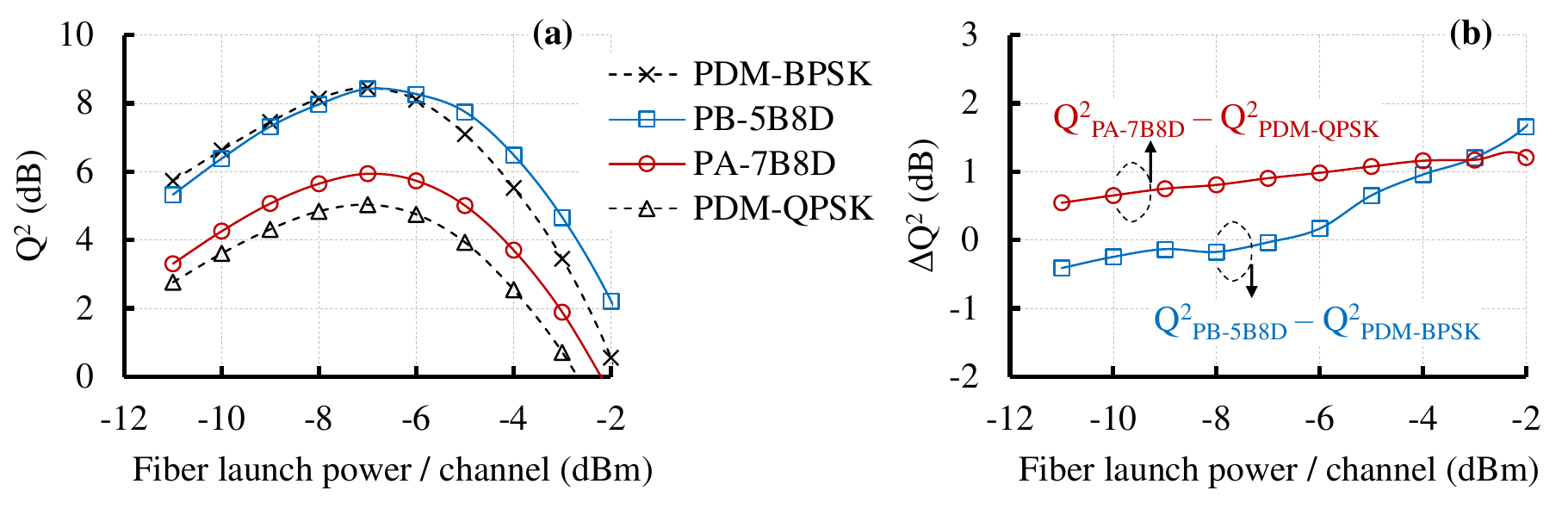}
	\caption{\label{figure1} Transmission system simulation results after 60spans for several values of the fiber launch power per channel, for PDM-BPSK, PB-5B8D, PA-7B8D and PDM-QPSK formats. (a) $Q^2$ factor. (b) Gain (expressed as $Q^2$ difference) of PB-5B8D and PA-7B8D over PDM-BPSK and PDM-QPSK, respectively.}
\end{figure*}
Figure \ref{figure1}.a shows the bell curves of PDM-BPSK, PB-5B8D, PA-7B8D and PDM-QPSK formats. In the noise-dominated linear regime, PDM-BPSK exhibits higher $Q^2$ factor values than PB-5B8D because of a higher minimum Euclidean distance ($Q^2$ factor higher by 0.4dB at -11dBm). Nevertheless, in the nonlinear regime, the PB-5B8D outperforms the PDM-BPSK, as shown in fig. \ref{figure1}.b. The increase of the $Q^2$ factor gain is proportional to the input power of the fiber and therefore to strength of fiber nonlinear impairments.\\
Comparing PA-7B8D to PDM-QPSK, fig. \ref{figure1}.a shows around 0.54dB linear performance difference at -11dBm launch power. As these two formats have the same minimum Euclidean distance, this difference comes from a lower bitrate for PA-7B8D. The difference increases slightly in the nonlinear regime as it can be seen in fig. \ref{figure1}.b (above 1dBm for minimum launch power of -5dBm), as PA-7B8D is more robust to nonlinear impairments.\\
\begin{figure*}[!!!ht]
	\centering
	\includegraphics[scale=0.85]{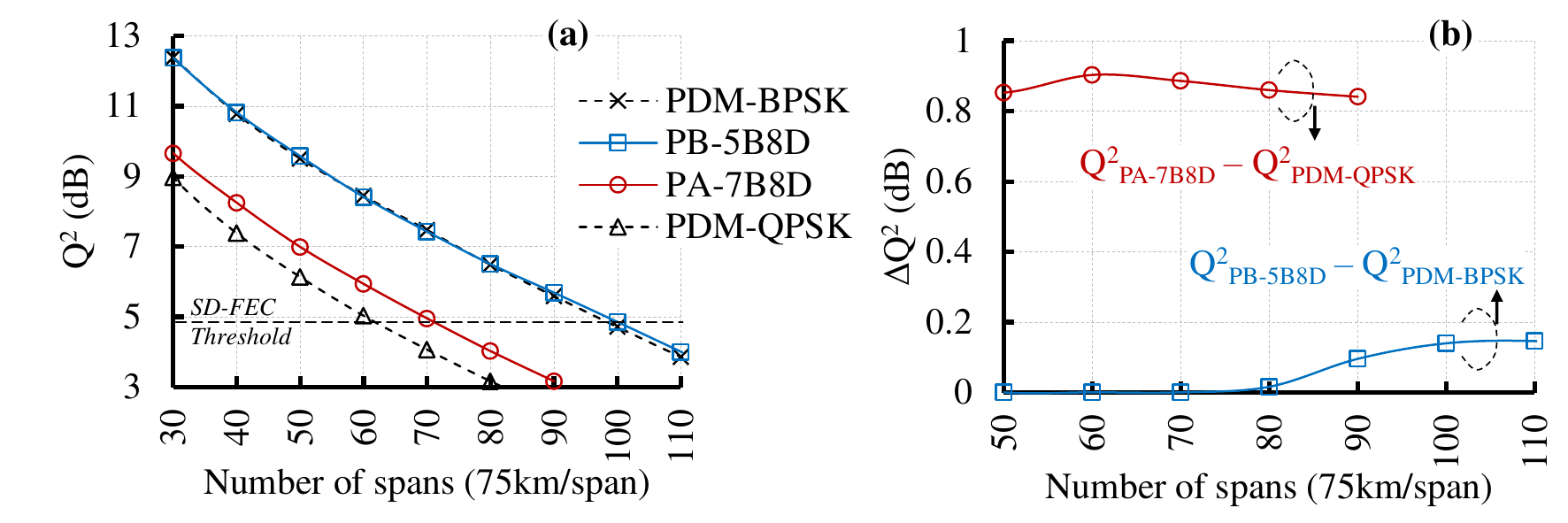}
	\caption{\label{figure2} Transmission system simulation results versus the number of spans, for PDM-BPSK, PB-5B8D, PA-7B8D and PDM-QPSK formats. (a) $Q^2$ factor. (b) Gain on $Q^2$ factor.}
\end{figure*}
After fixing the fiber launch power to its optimal value (-7dBm for all formats according to fig. \ref{figure1}.a), we simulate the reach dependence of each modulation format and plot the resulting $Q^2$ factors in fig. \ref{figure2}.\\
At a Soft-Decision Forward-Error-Correcting (SD-FEC) code threshold of $4.9$dB, we calculate the gain in reach of PB-5B8D over PDM-BPSK. A relatively small 1.6\% gain is achieved (fig. \ref{figure2}.a), however at a 25\% higher spectral efficiency. Despite a lower linear channel performance, PB-5B8D has still a longer reach. It can also be observed in fig. \ref{figure2}.b that this gain appears only after a transmission distance of 6750km. As a consequence, a significant gain in reach-capacity product is obtained for PB-5B8D with comparison to PDM-BPSK.\\
At the same SD-FEC threshold, a gain of 15\% in reach is achieved by PA-7B8D compared to PDM-QPSK baseline, at a 12.5\% lower  spectral efficiency (fig.\ref{figure2}.a). 
In fig.\ref{figure2}.b we observe a gain in $Q^2$ factor of at 0.9dB after 60 spans.  
Recalling that PA-7B8D has a gain of around 0.54dB in $Q^2$ factor over PDM-QPSK in the linear channel, therefore exhibits a higher tolerance to nonlinear impairments compared to PDM-QPSK. As a consequence, we find a slight increase of the reach-capacity product for PA-7B8D.\\
It is known that the PB constraint mitigates Cross Polarization Modulation (XPolM) effects~\cite{Shiner2014}. This is the reason why PB-5B8D performs better than PDM-BPSK in the nonlinear channel.
In simulations we have further observed that adding PA symbols for spectral efficiencies higher than 3bits/4D-symbol can still provide improved nonlinear performance compared to PDM-QPSK.
By avoiding strong XPolM-inducing PI symbols contained in PDM-QPSK, PA-7B8D provides a small net gain and a good trade-off between nonlinear performance and implementation complexity.
These features make our modulation formats useful in systems where XPolM is a limiting factor.

\section{Conclusion}
We have designed two new nonlinearity-tolerant 8D modulation formats at spectral efficiencies of 2.5 (PB-5B8D) and 3.5bits/4D-symbol (PA-7B8D). The PB-5B8D format outperforms standard PDM-BPSK by increasing the transmission reach, while offering a substantial 25\% increase in spectral efficiency. The PA-7B8D format can reach distances that are unreachable using PDM-QPSK. Even though the spectral efficiency is 12.5\% lower, it offers additional flexibility for the distance-capacity trade-off with a small net gain. We have provided a simple bit-to-symbol mapping, which shows that these format can be implemented with  slight modifications of the standard PDM-QPSK hardware. Finally, the two new modulation formats are very promising candidates for submarine transmission systems.

\bibliographystyle{ieeetr}
\bibliography{myrefs}

\end{document}